\documentclass[showpacs,amsmath,amssymb,aps,showkeys,floatfix,prd,a4paper]{revtex4}
\usepackage[dvips]{graphicx}
\usepackage{dcolumn}
\usepackage{bm} 
\usepackage{epsfig}
\usepackage{amsfonts}
\usepackage{amssymb,amscd}
\usepackage{subfigure}
\usepackage{xcolor}
\usepackage{verbatim}

\newcommand{\Pelotas}{High and Medium Energy Group, Instituto de F\'{\i}sica e Matem\'atica,
             Universidade Federal de Pelotas\\
             Caixa Postal 354,  96010-900, Pelotas, RS, Brazil.}
\newcommand{\UFRGS}{High Energy Physics Phenomenology Group, GFPAE  IF-UFRGS \\
Caixa Postal 15051, CEP 91501-970, Porto Alegre, RS, Brazil.}             
\newcommand{\IFUSP}{Instituto de F\'{\i}sica, Universidade de S\~{a}o Paulo,
           C.P. 66318,  05315-970 S\~{a}o Paulo, SP, Brazil.}

\begin{document}

\title{Investigating the exclusive vector meson photoproduction in nuclear collisions at Run 2 LHC energies}
\author{V. P. Gon\c{c}alves $^1$, M. V. T. Machado $^2$, B. D. Moreira $^1$, 
F. S. Navarra $^{3}$, G. Sampaio dos Santos $^1$}
\affiliation{$^{1}$ \Pelotas\\ $^{2}$ \UFRGS \\ $^{3}$ \IFUSP}

\begin{abstract}
We investigate the theoretical uncertainty on the predictions for the photoproduction of $\rho$ and $J/\Psi$
states in nucleus-nucleus collisions at Run 2 LHC energies using the Color Dipole formalism. The results for rapidity distributions and total cross sections are presented. 
Moreover, we compare directly the theoretical results to the recent preliminary data from ALICE collaboration on $\rho$ production in $PbPb$ collisions at central rapidity.
\end{abstract}


\maketitle

\section{Introduction}
The advent of the high-energy colliders has motivated the study of the hadron structure at high energies. In such scenario, a hadron becomes a dense system and the nonlinear effects inherent to the QCD dynamics may become visible. The proton structure can be studied through the photon-proton interaction, taking into account the QCD dynamics at high energies. The best place to study the hadron structure and QCD dynamics, especially with heavy nuclei, it will be the future electron-ion collider. Alternatively, one can study the $\gamma A$ interaction at the LHC, in ultraperipheral collisions (UPC). The main advantage of using colliding hadrons and
nuclear beams for studying photon induced interactions is the high photon-hadron center-of-mass energy and luminosities
achieved at LHC. Consequently, studies of $\gamma A$ interactions at the LHC could provide valuable information on the QCD dynamics at high energies. During the last years, the LHC has provided data on vector meson photoproduction
at Run 1 energies and in this year at Run 2 energies. The Run 2 at the LHC has already produced $PbPb$ collisions
and more data in hadronic collisions are expected in the next years. These collisions are now performed at energies which are a factor 2 larger than those of Run 1.  
Here, our goal is to analyze the theoretical uncertainties on the predictions for the photoproduction of
light and heavy vector mesons in $PbPb$ collisions at the LHC using the color dipole approach. 

\section{Formalism}
Let us start defining a UPC as a collision between two electric charges at impact 
parameters such that $b > R_{1} + R_{2}$, where $R_{i}$ is the radius of the charge 
$i$. In a UPC at high energies, it is well known that the hadrons act as a source of 
almost real photons. Consequently, the  exclusive meson photoproduction in hadron-hadron collisions can be factorized in terms of the equivalent flux of photons of the hadron projectile and photon-target production cross section \cite{upc},
\begin{eqnarray}
\sigma(h_{1} + h_{2} \rightarrow h_{1} \otimes V \otimes h_{2}) = 
\int d\omega \frac{n_{h_{1}}(\omega)}{\omega} 
\sigma_{\gamma h_{2} \rightarrow V \otimes h_{2}}\left( W_{\gamma h_{2}}^{2}\right) 
+ \int d\omega \frac{n_{h_{2}}(\omega)}{\omega} 
\sigma_{\gamma h_{1} \rightarrow V \otimes h_{1}}\left( W_{\gamma h_{1}}^{2}\right),
\label{epa}
\end{eqnarray}
where $n(\omega)$ is the equivalent photon spectrum and $\sigma_{\gamma h \rightarrow V \otimes h }( W_{\gamma h}^{2})$ 
is the vector meson photoproduction cross section. Moreover, the photon energy $\omega$ and 
$W_{\gamma h}$, the c.m. energy of the $\gamma h$ system, are related by $W_{\gamma h} = \sqrt{4 \omega E}$, where $E = \sqrt{s}/2$ 
and $\sqrt{s}$ is the hadron-hadron c.m. energy.
The equivalent photon spectrum is well known and is fully computed in QED. For the case where a 
nucleus is the source of photons \cite{upc},
\begin{eqnarray}
n_{A}(\omega) &=& \frac{2Z^{2}\alpha_{em}}{\pi }  
\left[\xi K_{0}(\xi) K_{1}(\xi) -\frac{\xi^{2}}{2} \left( K_{1}^{2}(\xi) - K_{0}^{2}(\xi)  
\right )  \right],
\label{epanuc}
\end{eqnarray}
where $\xi = \omega \left(R_{h_{1}} + R_{h_{2}} \right) / \gamma_{L}$,
and $\gamma_L$ is the Lorentz boost of a single beam.
The exclusive vector meson photoproduction in hadronic collisions can be evaluated using Eq.(\ref{epa}), which we need to know the  cross section for the process $\gamma \, h \to V \, h$,
\begin{eqnarray}
 \sigma(\gamma h \rightarrow V h) 
= \frac{1}{16 \pi}  \int_{-\infty}^{0} \left |{\cal A}^{\gamma h \rightarrow V h} (x, \Delta)  \right|^{2} dt.  
\label{cs_gammap}
\end{eqnarray}
Here, the $\gamma h$ interaction will be described within the dipole frame, where the
probing projectile fluctuates into a quark-antiquark pair with transverse separation $\textbf{\textit{r}}$ (and momentum fraction z) long after the interaction, which scatters off the hadron target and then forms a vector meson at the final state. If the lifetime of the dipole is much larger than the interaction time, a condition which is satisfied in high energy collisions, 
the scattering amplitude ${\cal A}^{\gamma h \rightarrow V h} (x, \Delta)$ can be written as \cite{KMW},
\begin{eqnarray}
{\cal A}^{\gamma h \rightarrow V h} (x, \Delta) =  
i \int dz \,\, d^{2}r \,\, d^{2}b_{h} [\Psi^{V*}\left(r, z \right)   
\Psi\left(r, z \right)]_T \,\,                              
e^{-i\left[b_{h} - (1-z) r \right]  \cdot \Delta}
\,\,2\,\, {\cal N}_{h}(x,r,b_{h}),
\label{amp}
\end{eqnarray}
where $\Psi^{V*}\left(r, z \right)$ and $\Psi\left(r, z \right)$ are the 
wave functions of the photon and of the vector meson, respectively. The overlap function 
$[\Psi^{V*} \left(r, z \right)\Psi\left(r, z \right)]_T$ describes the fluctuation of the photon with transverse polarization into a color dipole and the subsequent formation of the vector meson. All the information about the strong interactions in the process is encoded in the forward dipole-hadron 
scattering amplitude ${\cal N}_{h}(x,r,b_{h})$. The variable $b_{h}$ stands for the impact parameter, the separation
between the dipole center and the target center, $x=M_{V}^{2}/W^{2}$ is the Bjorken variable and $\Delta$ is related to the momentum transfer squared by $\Delta = \sqrt{-t}$.
In the numerical evaluations in next section, we have considered the Boosted Gaussian (BG) \cite{wfbg} and the
Light-Cone Gaussian (LCG) \cite{wflcg} wavefunctions and the
phenomenological models for dipole scattering amplitude: IIM \cite{IIM_plb}, bCGC \cite{Watt_bcgc,KMW} and IP-SAT \cite{ipsat1} models, which encode the main properties of the saturation approaches. 
The expressions for the overlap functions we have used
appropriately summed over the helicity and flavor indices
are given by
\begin{eqnarray}
(\Psi_{V}^* \Psi)_T = \hat{e}_f e \frac{N_c}{\pi z (1-z)}\left\{m_f^2K_0(\epsilon r)\phi_T(r,z) -[z^2+(1-z)^2]\epsilon K_1(\epsilon r) \partial_r \phi_T(r,z)\right\},
\end{eqnarray}
where $ \hat{e}_f $ is the effective charge of the vector meson, $m_f$ is the quark mass, $N_c = 3$, $\epsilon^2 = z(1-z)Q^2 + m_f^2$ and $\phi_T(r,z)$ define the scalar part of the  vector meson wave function. The BG and LCG models differ in the assumption about the function $\phi_T(r,z)$,
\begin{eqnarray}
\phi_T^{BG}(r,z) &=& N_T z(1-z) \exp\left(-\frac{m_fR^2}{8z(1-z)} - \frac{2z(1-z)r^2}{R^2} + \frac{m_f^2R^2}{2}\right), \\
\phi_T^{LCG}(r,z) &=& N_T [z(1-z)]^2 \exp\left(-\frac{r^2}{2R_T^2}\right).
\end{eqnarray}
The parameters $N_T$, $R$ and $R_T$ are  determined by the normalization condition of the wave function and by the decay width (see e.g. Refs.\cite{KMW,glauber1,bruno1,armesto_amir} for details).  
Several phenomenological models for the dipole scattering amplitude have been proposed in the literature, such models are 
based on the Color Glass Condensate formalism and describe the HERA data taking into account the 
nonlinear effects in the QCD dynamics. 
Three examples of very successful models are the  IIM, bCGC and IP-SAT models.    
The IIM model interpolates two analytical 
solutions of well known evolution equations (BFKL and BK equations), and ${\cal N}_{p}(x,r) $ is given by \cite{IIM_plb},
\begin{eqnarray}
{\cal N}_p(x,r) = \left\{  \begin{array}{l}
{\cal N}_{0}\left( 
\frac{rQ_{s}}{2}
\right )^{2[\gamma_{s}+(1/(\kappa \lambda Y))\ln (2/rQ_{s})]} 
\,\,\, , rQ_{s} \leq 2\\
1- e^{-A \ln^{2}(BrQ_{s})} \,\,\,\,\,\,\,\,\,\,\,\,\,\,\,\,\,\,\,\,\,\,\,\,\,\,\,\,\,\,\,\,\,\,\,\,\,\,\,\,\,\,\,, rQ_{s} > 2
\end{array} \right.  
\label{iim}
\end{eqnarray}
where $Y=\ln(1/x)$ and $Q_{s}(x) = \left(
x_{0}/x\right )^{\lambda/2}$ is the saturation scale.
In the bCGC model the form of ${\cal N}_p$ is the same as in 
Eq.(\ref{iim}), but the saturation scale has the following dependence on $b$
\begin{eqnarray}
Q_{s} \equiv Q_{s}(x,b) = \left(\frac{x_{0}}{x} \right )^{\lambda/2} \left[\exp \left(-\frac{b^{2}}{2B_{CGC}} \right) \right ]^{1/(2 \gamma_{s})}  .
\label{qs_bcgc}
\end{eqnarray}
We have considered the set of parameters
for the IIM and bCGC parameterizations from Ref.\cite{Rezaeian_update}. 
The IP-SAT model uses an eikonalized form for ${\cal N}_p$ that depends on a gluon distribution evolved via DGLAP equation and is written as 
\begin{eqnarray}
{\cal N}_p(x,r,b) = 1 - \exp \left[\frac{\pi^{2}r^{2}}{N_{c}} \alpha_{s}(\mu^{2}) \,\,xg\left(x, \frac{4}{r^{2}} + \mu_{0}^{2}\right)\,\, T_{G}(b) \right],
\label{ipsat}
\end{eqnarray}
with a Gaussian profile and the initial gluon distribution evaluated at $\mu_{0}^{2}$ are taken to be 
\begin{eqnarray}
T_{G}(b) &=& \frac{1}{2\pi B_{G}}  
\exp\left(-\frac{b^{2}}{2B_{G}} \right), \\
xg(x,\mu_{0}^{2}) &=&  A_{g}x^{-\lambda_{g}} (1-x)^{5.6}.
\end{eqnarray}
In this work we calculated ${\cal N}$ using a FORTRAN library provided by the authors of Ref.\cite{ipsat4}, which includes an updated analysis of combined HERA data.
In order to estimate $\gamma A$ process, we need to adapt the phenomenological models described above to the nuclear 
case. We will assume the model proposed in Ref.\cite{armesto} that contains the impact parameter dependence, where the dipole-nucleus scattering amplitude is written as
\begin{eqnarray}
{\cal N}_{A} &=& 1 - \exp \left[-\frac{1}{2} \sigma_{dip}(x,r^{2}) \, T_{A}(b_{A})
 \right], \label{Na_Glauber} \\
\sigma_{dip}(x,r^{2}) &=& 2 \int d^{2}b_{p} \,\, {\cal N}_{p} 
(x,r,b_{p}),
\label{eq14}
\end{eqnarray}
where $T_{A}(b_{A})$ is the nuclear thickness. The above equation  
sums up all the multiple elastic rescattering diagrams of the $q \overline{q}$ pair and is justified for large coherence length, where the transverse separation $r$ of partons in the multiparton Fock state of the photon becomes a conserved quantity, {\it i.e.} the size of the pair $r$ becomes eigenvalue of the scattering matrix.  

\section{Results}
\begin{figure}
\begin{tabular}{cc}
\centerline{
{
\includegraphics[height=6cm]{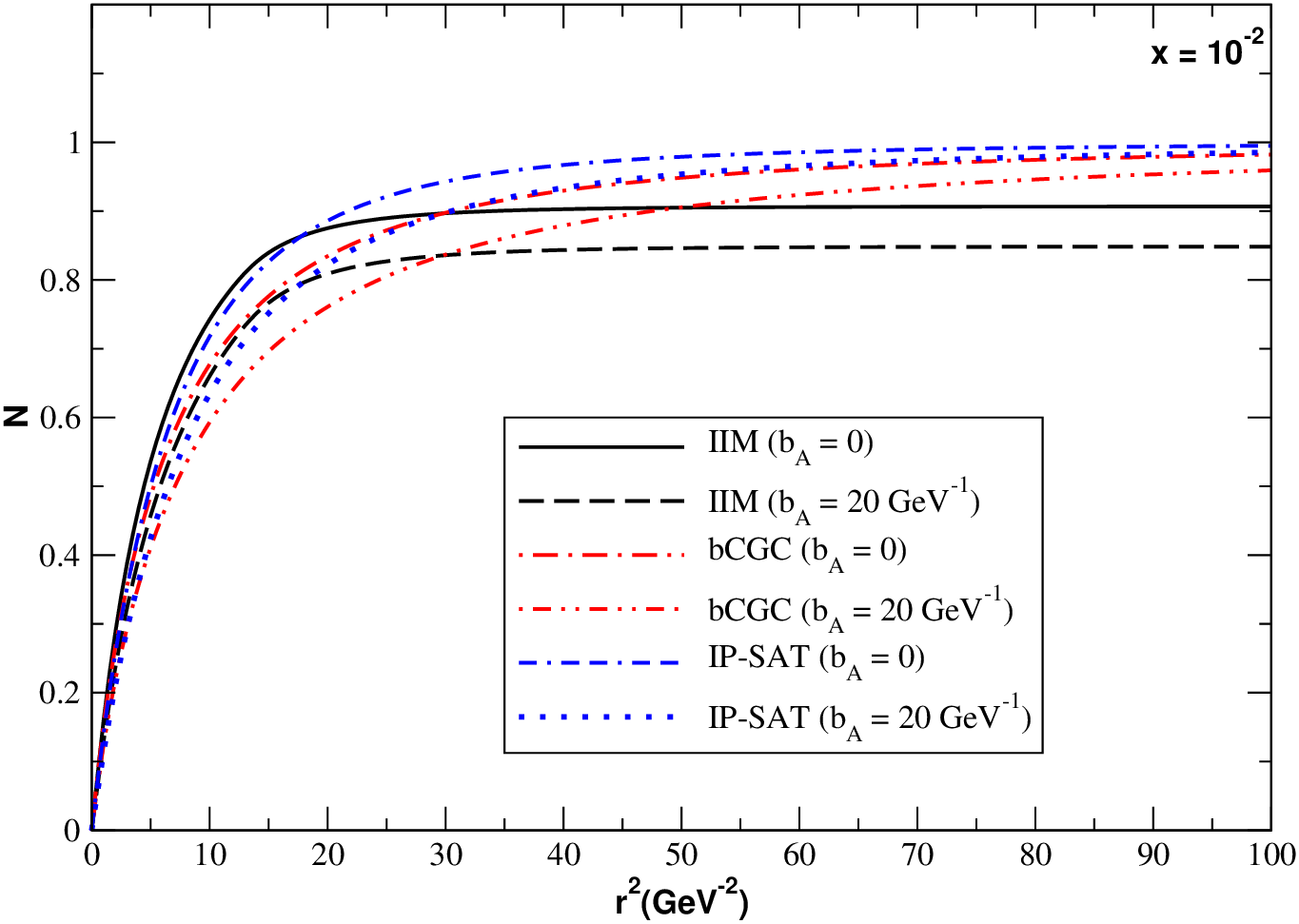}
}
{
\includegraphics[height=6cm]{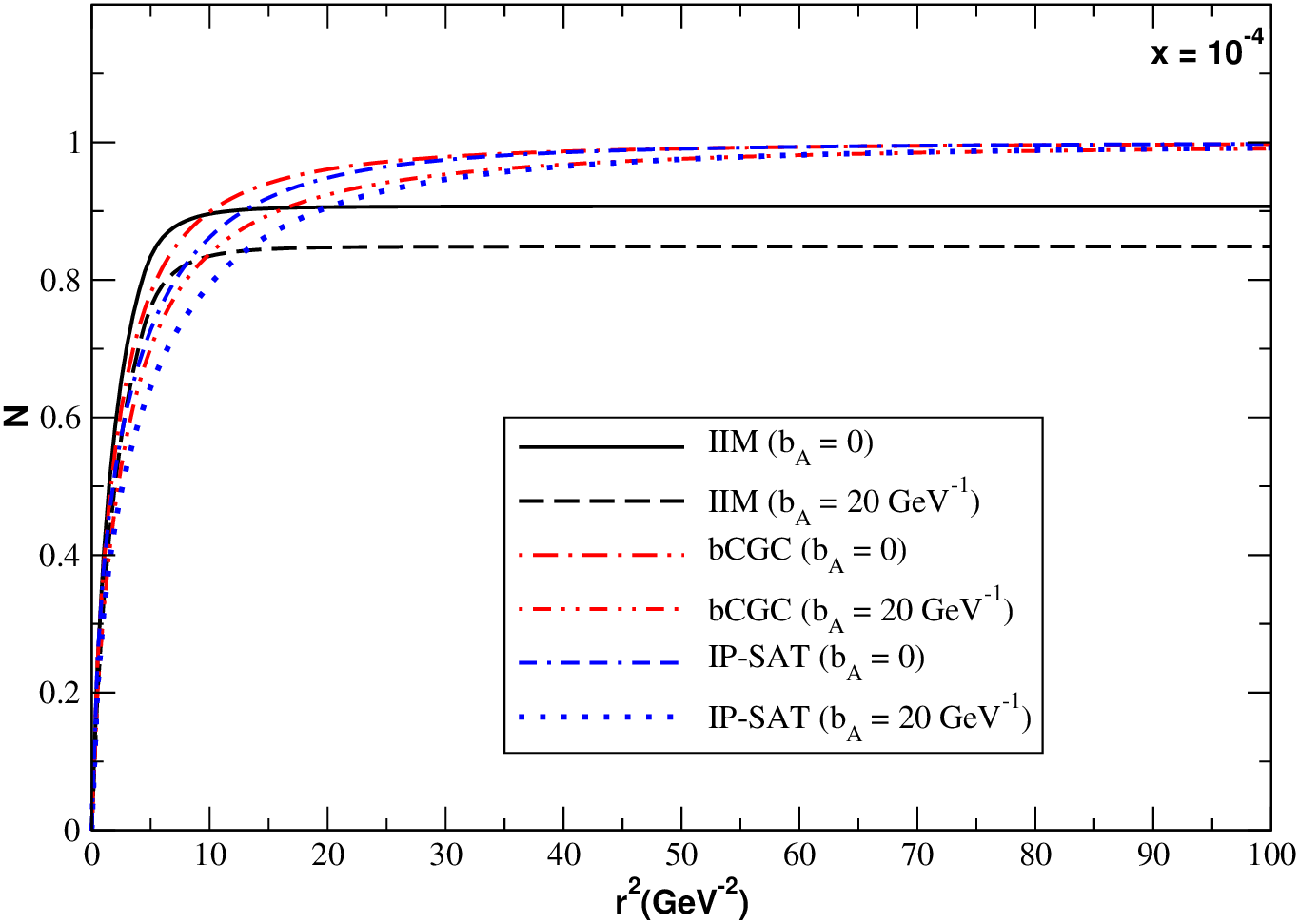}
}}
\end{tabular}
\caption{Dipole-nucleus scatering amplitude as a function of $r^2$ for fixed values of $b$ and different values of $x$ ($A=Pb$).}
\label{NA}
\end{figure}

Before presenting the predictions for the vector meson production we will compute $\mathcal{N}_A$ considering 
the different models for the dipole-proton scattering amplitude discussed previously.
In Fig.\ref{NA} we show the results for the nuclear scattering amplitude take into account the IIM, bCGC and IP-SAT models considering different values of the impact parameter $b_{A}$. As expected, $\mathcal{N}_A$ saturates faster for central collisions than for large impact parameters. Moreover, the deviation between the predictions are smaller.
This is directly associated with the model for ${\cal N}_{A}$, given by Eq.(\ref{Na_Glauber}), which is the same 
in all three cases. The future experimental data on vector meson photoproduction in $PbPb$ collisions will be useful to test this model of ${\cal N}_{A}$.

\begin{figure}
\begin{tabular}{cc}
\centerline{
{
\includegraphics[height=6cm]{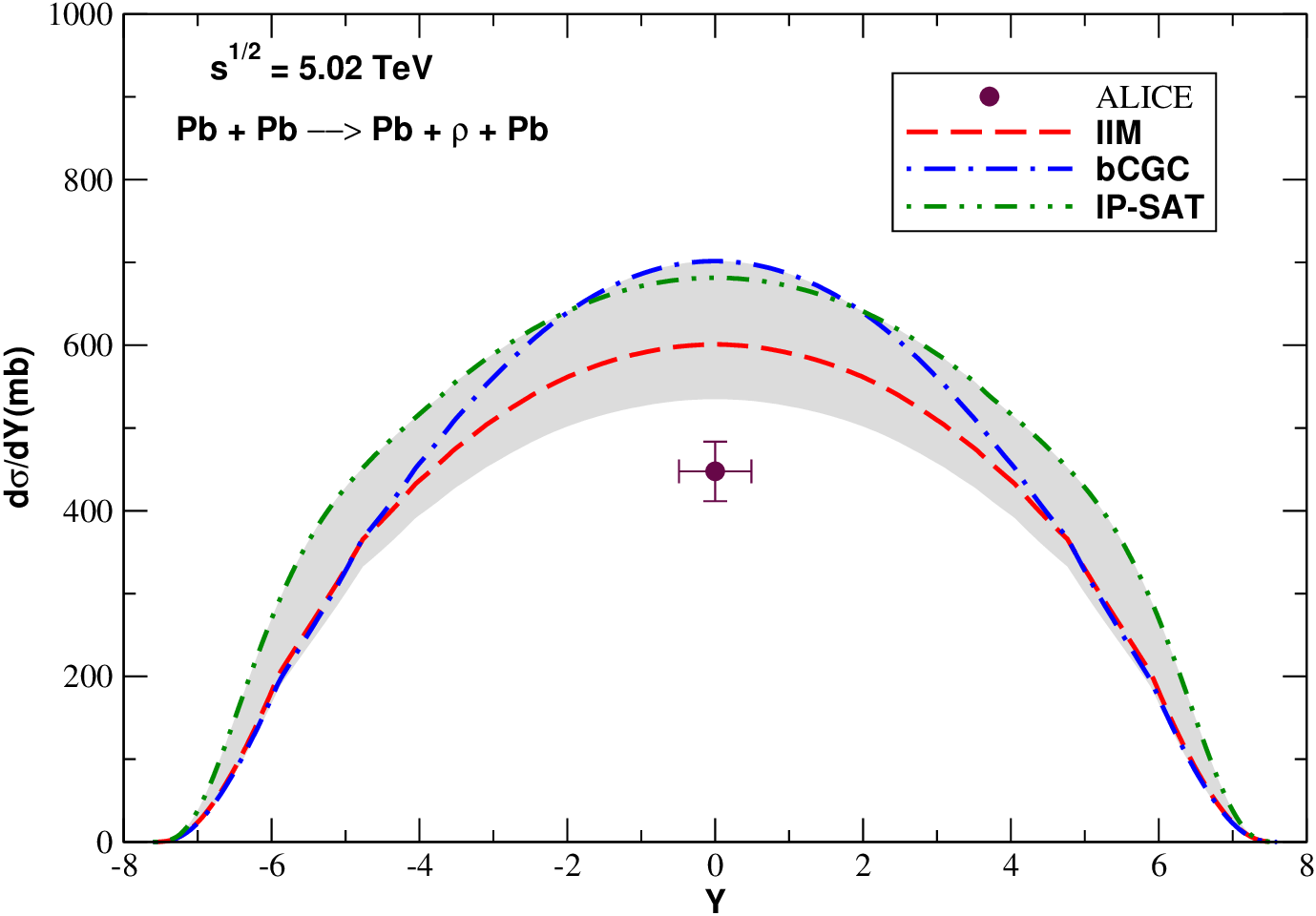}
}
{
\includegraphics[height=6cm]{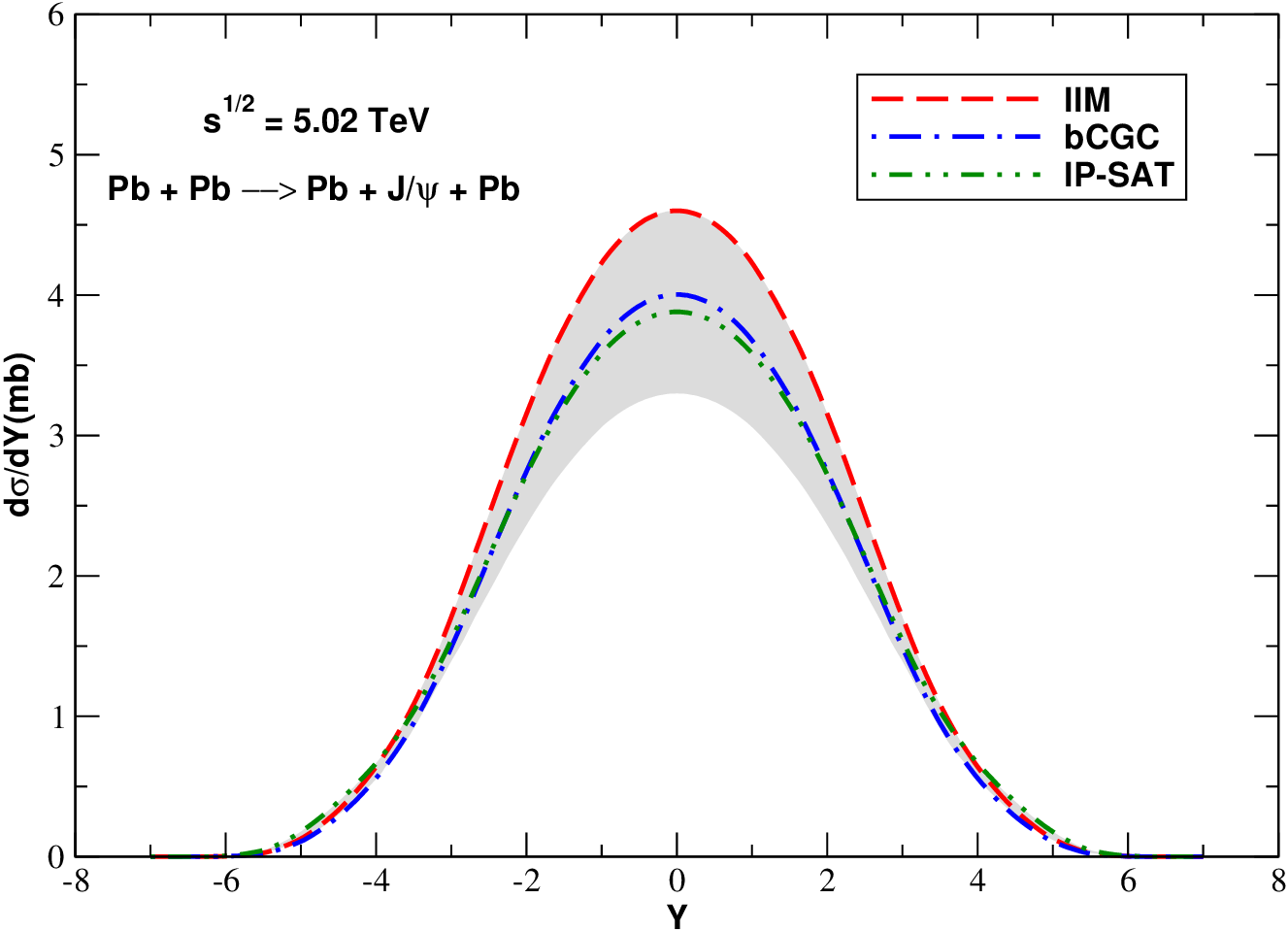}
}}
\end{tabular}
\caption{Rapidity distributions for the photoproduction of $\rho$ and $J/\Psi$ in $PbPb$ collisions at $\sqrt{s} = 5.02$ TeV.} 
\label{AA}
\end{figure}

In Fig.\ref{AA} we present our predictions \cite{Goncalves} (which are a complement the predictions presented in Ref.\cite{vicbruno}) for the exclusive photoproduction of $\rho$ and $J/\Psi$ in $PbPb$ collisions at $\sqrt{s} = 5.02$ TeV. As we can see the difference between the predictions is smaller as expected from Fig.\ref{NA}.  
In particular, at central rapidities the bCGC and IP-SAT parameteri\-zations deliver quite similar results. 
Moreover, in Fig.\ref{AA} we compare our results with the recent preliminary data on  
$\rho$ photoproduction at central rapidity released by ALICE Collaboration \cite{alice}. We can verify that the color dipole predictions overestimate the data. This result is an indication that other effects, not included 
in our analysis, should be included at least for this final state. Some possibilities are the inclusion of  
shadowing corrections \cite{glauber1} or absorption corrections 
\cite{Martin}. Another possible conclusion is that the treatment of 
the dipole-nucleus interaction, described here by the model presented 
in Eq.(\ref{Na_Glauber}), should be improved. 

Finally, in order to estimate the dependence of our results on the models used (for the dipole scattering amplitude and for the vector meson wavefunctions), we present 
in Tab.\ref{sec_tot_tab} the lower and upper bounds of our predictions for the total cross sections. As expected from the analysis of the rapidity distributions, the largest uncertainties are presented in the $\rho$ production. Additionally, the cross sections for $PbPb$ collisions decrease with the mass of the vector meson. 

\begin{table}
\centering
\begin{tabular}{|c|c|c|}\hline
 & $\rho$                  & $J/\Psi$         \\ \hline
PbPb ($\sqrt{s} = 5.02$ TeV) &   5.26 - 7.04 b    &18.24 - 24.47 mb      \\ \hline
\end{tabular} 
\caption{The total cross sections of the photoproduction of $\rho$ and $J/\Psi$ in $PbPb$ collisions at the Run 2 LHC energies.}
\label{sec_tot_tab}
\end{table}

\section{Summary}
We have presented predictions for the exclusive photoproduction of $\rho$ and $J/\Psi$ 
in $PbPb$ collisions. 
Our results demonstrated that cross section for the light meson production is larger than the heavy meson one. Although the light meson production suffers from the largest theoretical uncertainties on the predictions.
The comparison with the preliminary experimental  
data on the $\rho $ production in $Pb Pb$ collisions indicated that the current color 
dipole description overestimate the data. 
This can be interpreted as an indication that a more careful 
treatment of the dipole-nucleus interaction and/or another effects as shadowing and absorptive corrections should be incorporated to the formalism. 

\begin{acknowledgments}
This work was  partially financed by the Brazilian funding agencies CAPES, CNPq, FAPERGS and FAPESP. 
\end{acknowledgments}

\hspace{1.0cm}

\end{document}